\documentstyle[aps,prb]{revtex}

\renewcommand{\ref}[1]{\raisebox{.6ex}{[#1]}}

\newcommand{\be}{\begin{equation}}
\newcommand{\ee}{\end{equation}}

\newcommand{\ba}{\begin{array}}
\newcommand{\ea}{\end{array}}

\begin{document}

\title{ Friction Coefficient and Berry Phase for 
  a Topological Singularity in a 
         Superfluid }

\author{X.-M. Zhu}

\address{ Department of Experimental Physics, 
    Ume\aa{\ }University, 901 87 Ume\aa, SWEDEN } 

\author{ P. Ao}
\address{ Department of 
   Theoretical Physics, Ume\aa{\ }University, 901 87 Ume\aa, SWEDEN}

\maketitle
       
\abstract{
   The coexistence of the universal transverse force and the non-universal 
   friction force on a topological singularity in a superfluid is shown 
   here.  Based on the BCS type microscopic theory, 
   we explicitly evaluate the quasiparticle contribution to the friction 
   coefficient in a clean fermionic superfluid,
   showing a new feature of logarithmic divergence.}



\section{INTRODUCTION}

When a vortex in a superfluid moves,
in addition to the hydrodynamic Magnus force
which is proportional to the superfluid density,  
the coupling of the 
vortex to quasiparticles, phonons and impurities may
cause extra forces
in both transverse and longitudinal directions of the vortex motion.
Those possible non-vanishing forces
have been the underlying assumption for phenomenological models
used in analyzing experiments\cite{donnelly}, though analyses based on 
topological methods such as the Berry phase always give
the universal transverse force.\cite{at}

While the total transverse force on a moving vortex needs further 
investigations,
the friction force is poorly studied. The usual quoted friction
coefficient formula derived from a microscopic Hamiltonian is 
obtained through a
relaxation time approximation in force-force correlation 
functions.\cite{kopnin} 
However, this type of procedure is incorrect\cite{ao},
for the reason known since the 60's in the context of
obtaining the friction of electrons using force-force correlation functions. 
The lack of an explicit and correct
microscopic derivation of the friction has two consequences.
First, though the friction of vortex motion is
an important quantity in many experiments, it has not been received
due attention. Second,
 the incomplete understanding of the friction has generated doubts 
about the exact result on the transverse force. 

Here we describe 
a self-contained theory of both the friction and the transverse 
forces by a rigorous and elementary method.
The method we choose follows that of Ref.[5].
In order to obtain the friction,
we use a limiting process
similar to the one in transport theory 
 and in non-equilibrium statistical mechanics, 
which  has not been explicitly discussed in Ref.[5].

In the following, after giving a general
expression for the friction in Section 2, as an example,
we  evaluate the
friction for the case of a clean fermionic superfluid
in Section 3.
The resulting friction coefficient is new.
It comes 
from the off-diagonal potential scattering of the extended quasiparticles, 
and is stronger than the ohmic damping by a logarithmic diverging factor.

\section{ TRANSVERSE AND FRICTION FORCES }

We consider an isolated vortex in the superfluid whose position ${\bf r}_0$ 
is specified by a pinning potential.
The system is otherwise homogeneous and infinite.
There is no externally applied supercurrent or normal current.
The vortex is allowed to move slowly. Hence the system Hamiltonian $H$
contains a slowly varying parameter ${\bf r}_0(t)$. The many-body
wavefunction of the superfluid $| \Psi_\alpha(t) \rangle $
can be expanded in terms
of the instantaneous eigenvalues $E_\alpha({\bf r}_0 )$ and eigenstates
$|\psi_\alpha( {\bf r}_0 )\rangle $, for which we choose phases such that
$\langle\psi_\alpha|\dot\psi_\alpha \rangle=0$. Because of our
assumption of a  homogeneous superfluid, 
 $E_\alpha({\bf r}_0 )$ is independent of both ${\bf r}_0 $ and  time $t$.
With those considerations, the many-body wavefunction 
$| \Psi_\alpha(t) \rangle $ can be expressed by
\be
   |\Psi_\alpha(t)\rangle   =  e^{-iE_\alpha t/\hbar}
   |\psi_\alpha( {\bf r}_0 ) \rangle +\sum_{{\alpha'}\ne\alpha}
     a_{\alpha'}(t)
     e^{-i E_{\alpha'} t/\hbar} |\psi_{\alpha'}({\bf r}_0)\rangle \;, 
\ee
where, to the first order in velocity, $a_\alpha(t)=1$, and
\be
  a_{\alpha'}(t)= - \int^t_0 dt' \langle\psi_{\alpha'}|\dot\psi_\alpha \rangle
   e^{ i(E_{\alpha'} - E_\alpha ) t'/\hbar} \;.
\ee
This gives the
expectation value of the force on the vortex as
\begin{eqnarray}
  {\bf F} & = & -\sum_\alpha f_\alpha \langle\Psi_\alpha|
   \nabla_0H|\Psi_\alpha \rangle  \nonumber  \\
      &  =  &
  -\sum_\alpha f_\alpha \langle\psi_\alpha|
   \nabla_0H|\psi_\alpha \rangle
 + \sum_{{\alpha'}\ne\alpha}f_\alpha 
    \langle\psi_\alpha({\bf r}_0(t) )|
    \nabla_0H|\psi_{\alpha'}({\bf r}_0(t) )\rangle \times \nonumber \\
 & &   \int^t_0 dt'  \langle\psi_{\alpha'}( {\bf r}_0(t') )|
   \dot{\psi}_\alpha({\bf r}_0(t') )\rangle 
  e^{ i(E_{\alpha'} - E_\alpha )(t'-t)/\hbar }     + c.c. \; , 
\end{eqnarray}
where $f_{\alpha}$ is the occupation probability of the state ${\alpha}$.
    For a vortex moving with a small and uniform velocity 
${\bf v}_V = \dot{\bf r}_0(t) $,
\be
  |\dot{\psi}_\alpha( {\bf r}_0  )\rangle 
  ={\bf v}_V\cdot | \nabla_0 \psi_\alpha(  {\bf r}_0 )\rangle \; . 
\ee
Here $\nabla_0$ denotes the partial derivative with respect to
the position ${\bf r}_0$ of the pinning potential, the vortex position.
The first term in the right hand side of Eq.(3)  is
independent of ${\bf v}_V$ and will be ignored.
In fact, it is zero because of the translational invariance regarding the 
vortex position.
The integration in time
can be carried out directly. In addition, we also need to use  the following
relations, 
\be
   \langle\psi_\alpha|\nabla_0H|\psi_{\alpha'} \rangle
   = (E_{\alpha'}- E_\alpha )
    \langle \psi_\alpha|\nabla_0\psi_{\alpha'} \rangle 
   = (E_\alpha - E_{\alpha'} )
    \langle \nabla_0 \psi_\alpha| \psi_{\alpha'} \rangle  \; , 
\ee
which are obtained by taking gradient $ \nabla_0$ with respect to
$ H |\psi_{\alpha'} \rangle =  E_{\alpha'} |\psi_{\alpha'} \rangle$
and $   \langle \psi_\alpha|H =  E_\alpha \langle \psi_\alpha|$,
then multiplying  from left or right by 
$ \langle \psi_\alpha|$ or $ |\psi_{\alpha'} \rangle $ respectively.

To obtain the long time behavior 
of Eq.(3) requires a limiting procedure. There are two ways of
taking limiting sequences. A possible sequence is to take
the low frequency limit before the thermodynamic limit. 
As we will find out, the transverse force is independent of the
limiting process. Therefore such a limiting sequence is correct as long
as only the transverse force is concerned. In such a calculation, because 
the energy levels
have been treated as discrete ones, there is no friction.
In order to obtain friction,
we have to take the thermodynamic limit before the low frequency limit.
For this purpose, we use the Laplace average\cite{kohn},
 $ \lim_{t\rightarrow\infty}{\bf F}(t)
  = \lim_{\epsilon\rightarrow 0^+}
   {\bf F}(\epsilon) $, where
$ {\bf F}(\epsilon)= \epsilon \int_{0}^{\infty} dt \; 
          {\bf F}(t) e^{-\epsilon t}$, and
  ${\bf F}(\epsilon = 0) = {\bf F}_{\perp} + {\bf F}_{\parallel} $. 
We will also use the identity $
  1/( \epsilon + i x)  =
  \pi\delta(x)-i P ( 1 /x ) $. 
For the transverse force ${\bf F}_{\perp}$, we have
\[ 
  \lim_{\epsilon\rightarrow 0^+}\epsilon
  \int_{0}^{\infty} dt  e^{ - \epsilon t} 
  \left(1-\cos(E_\alpha- E_{\alpha'})t/\hbar\right)
  = \lim_{\epsilon\rightarrow 0^+}
  \left(1 - \pi\delta( E_\alpha- E_{\alpha'} ) \epsilon \right) = 1\; ,
\]
because the summation over states in Eq.(3) is well behaved
regardless whether $ E_\alpha$ is discrete or continuous.
Hence
\be
  {\bf F}_{\perp} = i \hbar \sum_{{\alpha'}\ne\alpha}
     f_\alpha \; \left\{ \left(
     \langle\psi_\alpha|\nabla_0\psi_{\alpha' } \rangle
    \times \langle \nabla_0\psi_{\alpha'}|\psi_\alpha\rangle \right)
     \cdot\hat{\bf z}\right\} \; 
      {\bf v}_V \times \hat{\bf z} \; .
\ee 
This is precisely what has been obtained in Ref.[5], where 
further calculations lead to
 $ {\bf F}_{\perp} = - h L\rho_{s} {\bf v}_V \times \hat{\bf z} $,
independent of details of the system. 
Here $\rho_s$ is the superfluid number density, $L$ the length of the vortex. 
The universal nature of the transverse force
has been confirmed in experiments of the vibrating wire\cite{vinen} and 
the vortex precession\cite{zieve}, and even in dirty 
superconductors\cite{zhu}. 
The longitudinal force, friction, is given by 
${\bf F}_{\parallel}=-\eta  {\bf v}_V$ with
\be
 \eta = \frac{\pi }{2}
    \sum_{{\alpha'}\ne\alpha }   \hbar  
     \frac{f_\alpha - f_{\alpha'} }{ E_{\alpha'}- E_\alpha }
  \delta( E_\alpha- E_{\alpha'} )
   |\langle\psi_\alpha|\nabla_0H|\psi_{\alpha' } \rangle |^2  \, .
\ee
The friction coefficient $\eta$ is determined by low energy 
excitations such as phonons, extended quasiparticles,
and localized quasiparticles when their discrete 
energy spectrum is smeared out by impurities. 
This expression 
is identical to the result in Ref.[10] for the case 
of ohmic damping in the zero frequency limit. 
Eq.(7) will not pick up any superohmic contributions, 
and will give infinity for any subohmic contributions.
 We point out that with the aid of Eq.(5)
the transverse force can be expressed only in terms of the wavefunction 
or the density matrix without explicit referring to the Hamiltonian or 
its eigenvalues, as shown by Eq.(6), on the other hand,
the longitudinal force, the friction, cannot, as shown by Eq.(7).
The explicit dependence on the Hamiltonian or its eigenvalues in Eq.(7) 
is the source of the sensitivity of friction to details of the system.

\section{ QUASIPARTICLE CONTRIBUTION  }

For the superfluid $^4$He, there is no full microscopic theory yet
\cite{donnelly}.
We will not attempt to evaluate Eq.(7) for this superfluid here
because of the sensitivity of the friction to details.
The situation can be very different 
in the case of  the recent Bose-Einstein condensed systems, where
a well defined microscopic theory is supposed to be known.
Instead, as an example to illustrate the directness and usefulness, 
we evaluate Eq.(7) 
for the case of a homogeneous fermionic superfluid
using BCS theory with s-wave pairing. 
At finite temperatures the extended states above the Fermi level,
the quasiparticles, are partially occupied. 
The vortex motion causes transitions between these states and gives rise to
friction.
The transitions between different single quasiparticle levels
$ \langle\psi_\alpha|\nabla_0H|\psi_{\alpha'} \rangle$
are considered here since they dominate the low energy process.
The quasiparticles
are described by the eigenstates, $u_\alpha $ and $v_\alpha $,
of the Bogoliubov-de Gennes equation. Their behavior in the 
presence of a vortex has been well studied in Ref.[11]. 
We may take
\be
   |\psi_\alpha \rangle  = 
   \left( \begin{array}{c} u_\alpha (x) \\ v_\alpha (x) \end{array} 
       \right) = \frac{1}{\sqrt{L} }
    e^{ik_z z}e^{i\mu\theta + i\sigma_z\theta /2}\hat{f}(r) \; ,
\ee
with ${\bf r}$ measured from the vortex position, and $\theta$ the azimuthal
angle around the vortex.
In order to obtain an analytical form for the transition element,
we use a WKB solution for $\hat{f}(r)$,
\begin{eqnarray}
 \hat{f}(r)& = &  \frac{1}{ 2 \sqrt{ R\; r } }
   \left( \begin{array}{c} 
      [ 1 \pm \sqrt{ E^2 - |\Delta (r) |^2}/E ]^{1/2}  \\ { }
     [ 1 \mp \sqrt{E^2- |\Delta (r)|^2}/E ]^{1/2}   \end{array} 
       \right)\times \nonumber \\
 & &  \exp{\left\{i\int_{r_t}^r dr' \left( k_{\rho}^2 
      \frac{r'^2-r_t^2}{r'^2}\pm \frac{2m}{\hbar^2 } 
    \sqrt{ E^2- |\Delta (r')|^2}\right)^{\frac{1}{2} } \right\}} + c.c.  \; .
\end{eqnarray}
Here $k_{\rho}^2 = k_f^2 - k_z^2$,  $R$ is the radial size of the
system. This WKB solution 
is valid when $r$ is outside the classical turning point
$r_t = |\mu|/k_{\rho}$. 
Here $r_t $ is the impact parameter.
A WKB solution also exists inside
the turning point. However, because it approaches zero as $ 
(r k_{\rho})^{|\mu|} /|\mu|! $, the contribution to the transition elements
from this region is small, 
and will be set to zero.  
The transition elements are then given by
\begin{eqnarray}
   |\langle\psi_\alpha|\nabla_0H|\psi_{\alpha' } \rangle|^2
   & = & \left|\int dx ( u_{\alpha'}^{\ast}(x) \nabla_0 \Delta v_\alpha(x) + 
         v_{\alpha'}^{\ast}(x) \nabla_0 \Delta^{\ast} u_\alpha(x) ) \right|^2  
   \nonumber \\
   & = &  \left\{  \begin{array}{lc} 
    \frac{\Delta_\infty^4}{E^2R^2} \; 
    \delta_{k_{z1},k_{z2}} \delta_{\mu_1 ,\mu_2\pm1}
   \, , \; & |\mu| \leq \xi_0 k_{\rho} \\
   0 \, , & |\mu| >\xi_0 k_{\rho} \,  \end{array} \right. \; .
\end{eqnarray}
Here $\Delta_\infty$ is the value of $|\Delta (r)|$ 
far away from the vortex core.
Physically it means that if the classical quasiparticle trajectory 
is far away from
the vortex core, it will not contribute to the friction.
The summation over states in Eq.(7) is replaced by 
\[
  \sum_{{\alpha'}\ne\alpha}=
  \sum_{\mu_1 , \mu_2 , k_{z1}, k_{z2}} 
  \int dE_1 dE_2 
  \frac{E_1}{\sqrt{E_1^2-\Delta_\infty^2}}
  \frac{E_2}{\sqrt{E_2^2-\Delta_\infty^2}}
  \left(\frac{m}{\hbar^2 k_f}\right)^2\frac{R^2}{\pi^2} \, ,
\]
after considering the density of states. 

Substituting Eq.(10) into Eq.(7), using the quasiparticle distribution 
function $f_\alpha  = 1/(e^{\beta E_\alpha} + 1 )$, 
 the coefficient of friction is given by
\be
  \eta =
   \frac{Lm^2\xi_0 \Delta_\infty^4\beta}{8\pi^2 \hbar^3} 
   \int_{\Delta_0}^{\infty} dE 
   \frac{E^2}{E^2-\Delta_\infty^2}
   \frac{1}{E^2\cosh^2{(\beta E/2)}} \;  .
\ee
The integral in Eq.(11) diverges logarithmically. 
It implies that the spectral function
corresponding to the vortex-quasiparticle coupling is not
strictly ohmic but has an extra frequency factor proportional to
$\ln(\Delta_\infty/\hbar\omega )$\cite{cl,az}.
When  $\hbar\omega$ is not very small comparing to $\Delta_\infty$, 
which may be realized   when  close to T$_c$,
we can ignore the logarithmic divergence 
in Eq.(11) by using the density of states
for normal electrons to obtain a finite friction,
i.e. replacing  $E^2/(E^2-\Delta_\infty^2)$ with 1 in Eq.(11).
Close to  T$_c$, the friction approaches zero the same way as 
$\Delta_\infty^2$, which is proportional to the superfluid density $\rho_s$.
When $ - \ln(\hbar\omega/\Delta_\infty)$ is large, we
need to use a more accurate
expression of vortex friction obtained in Ref.[10]. 
Straightforward evaluation shows that in such a case
\be
   \eta = \frac{Lm^2\xi_0\Delta_\infty^3\beta}{16\pi^2\hbar^3 }
   \frac{1}{\cosh^2{(\beta \Delta_\infty /2)}}
   \ln (\Delta_\infty/ \hbar\omega_c ) \, .
\ee
Here $\omega_c$ is the low frequency cut-off.
It is determined by the size of the system for a single vortex, 
and by the inter-vortex distance for a vortex array.

It should be emphasized that the logarithmic
divergence comes from the interplay between  the 
divergence in the density of states and the off-diagonal potential scattering.
We can consider a situation in which we physically create a pinning center to
trap the  vortex and guide its motion. In such a case the vortex
has a diagonal potential. 
If the scattering is dominated by the diagonal potential, e.g., by the
pinning potential, an additional factor coming from $|u_\alpha|^2 
- |v_\alpha|^2$ will remove this logarithmic divergence.
This again shows the sensitivity of the friction to system details.

The above results
apply to the dynamics of the axisymmetric vortex 
in  $^3$He  B phase, where the 
Bogoliubov-de Gennes equation is essentially the same as the 
one in a s-wave superconductor. 

\section{ CONCLUSION }

The transverse force formula confirms  the
Berry phase results obtained with a trial many-body wavefunction.
New results on the friction have been obtained from quasiparticle
contributions, which can be further tested 
experimentally.

\section*{ACKNOWLEDGMENT}  

{ This work was financially supported by Swedish NFR. }


\begin{thebibliography}{99}

\bibitem{donnelly}
 R.J. Donnelly, {\it Quantized Vortices in Helium II}, Cambridge University
  Press, Cambridge, 1991.
\bibitem{at}
   P. Ao and D.J. Thouless, {\it Phys. Rev. Lett.} {\bf 70}, 2158 (1993); 

   F. Gaitan, {\it Phys. Rev.} {\bf B51}, 9061 (1995). 
\bibitem{kopnin}
   N.P. Kopnin, {\it Physica} {\bf B210}, 267 (1995). 
\bibitem{ao}
   P. Ao, {\it Phys. Rev. Lett.} {\bf 80}, 5025 (1998).
\bibitem{tan}
   D.J. Thouless, P. Ao, and Q. Niu, 
          {\it Phys. Rev. Lett.} {\bf 76}, 3758 (1996).
\bibitem{kohn}
  W. Kohn and J.M. Luttinger, {\it Phys. Rev.} {\bf 108}, 590 (1957). 
\bibitem{vinen}
  W.F. Vinen, {\it Proc. Roy. Soc. (London)} {\bf A 260}, 218 (1961);
 
  S.C. Whitmore and W. Zimmermann Jr., 
           {\it Phys. Rev.} {\bf 166}, 181 (1968). 
\bibitem{zieve}
  R.J. Zieve {\it et al.},
   {\it J. Low Temp. Phys.} {\bf 91}, 315 (1993).
\bibitem{zhu}
  X.-M. Zhu, E. Brandstrom, and B. Sundqvist, 
      {\it Phys. Rev. Lett.} {\bf 78}, 122 (1997).
\bibitem{az}
   P. Ao and X.-M. Zhu, {\it Physica} {\bf C282-287},  367 (1997).     
\bibitem{bardeen}
 J. Bardeen {\it et al.},
    {\it Phys. Rev.} {\bf 187}, 556 (1969). 
\bibitem{cl}
  A.O. Caldeira and A.J. Leggett, 
   {\it Ann. Phys.(NY)} {\bf 149}, 374 (1983). 
%
%
\end{thebibliography}
\end{document}